\documentclass[showpacs,preprintnumbers,amsmath,amssymb,aps, prd]{revtex4}
\usepackage{graphicx}
\usepackage{grffile}
\usepackage{mathrsfs, mathtools}
\usepackage{caption}
\usepackage{float}
\usepackage{xcolor}
\usepackage{dcolumn}
\usepackage{bm}
\usepackage{graphicx,subfigure,epsfig}
\usepackage{multirow}

\begin{document}
\title{Study of quasinormal modes, greybody bounds, and sparsity of Hawking radiation
within the metric-affine bumblebee gravity framework}
\author{Sohan Kumar Jha}
\email{sohan00slg@gmail.com}
\affiliation{Chandernagore College, Chandernagore, Hooghly, West
Bengal, India}
\author{Anisur Rahaman}
\email{anisur.associates@iucaa.ac.in; manisurn@gmail.com
(Corresponding Author)} \affiliation{Durgapur Government College,
Durgapur, Burdwan - 713214, West Bengal, India}

\date{\today}
\begin{abstract}
\begin{center}
Abstract
\end{center}
We consider a static and spherically symmetric black hole metric
that emerges from the vacuum solution of the traceless
metric-affine bumblebee model. Our study focuses on the possible
implications of the modifications induced by the model on various
astrophysical observables that include quasinormal modes, ringdown
waveforms, Hawking radiation spectrum, sparsity of that radiation,
and the lifetime of a black hole. We explore the impact of the
Lorentz symmetry-breaking parameter $\alpha$ on the quasinormal
modes with the help of the $6th$-order WKB method. Our inquisition
reveals that the emission frequency and decay rate initially
decrease with $\alpha$ and then grow up. As a result, the LSB
becomes critically important for maintaining the stability of the
system after being exposed to perturbation. The convergence of the
WKB method for various orders is also studied here. We then
analyze the Hawking temperature, radiation spectrum, and sparsity
in this modified gravity framework that provides valuable insights
into the thermal radiation emitted by black holes. It points out
that the Hawking temperature, the peak of the power spectrum, and
the total power emitted initially decreases and then increases with
$\alpha$. However, The variation of the sparsity with $\alpha$
follows a reverse trend. Finally, we obtain the analytical
expression of the 'lifetime' of black holes and scrutinize the
effect of $\alpha$ on it.

\end{abstract}
\maketitle
\section{Introduction}
Black hole perturbation theory plays a prominent role in the study
of quasinormal modes (QNMs). Regge and Wheeler conducted an
innovative work on perturbation around black holes \cite{REGE},
and that pioneering study provided the foundation for a plethora
of other significant works. A black hole oscillates with complex
frequencies at the intermediate stage when it is subjected to
non-radial perturbation. These oscillations are referred to as
QNMs. Press \cite{PRESS} first used the term QNMs, but
Vishveshwara \cite{VISH} initially identified them in the
simulations of gravitational wave scattering off a Schwarzschild
black hole. For a perturbed black hole, the QNMs are the
frequencies of oscillation that crucially depend on the
characteristics of black holes like mass, spin, and charge
\cite{KOKKO, HPN, RKONO, CARDOSO}. These modes are identified by a
collection of discrete (often incomplete) complex frequencies.
Information obtained from gravitational waves combined with
findings in electromagnetic spectra \cite{LBAR, KAKI, GODDI}
should one day enable us to determine the fundamental parameters
(mass, angular momentum, and charge) of the black holes, and that
in turn, will allow the testing of the theory of gravity at the
strong field limit. However, there is currently a great deal of
uncertainty regarding the precise determination of these
fundamental parameters of black holes, which leaves a lot of space
for alternative or modified theories of gravity \cite{KONO}. In
this contest, along with the other modified theories, bumblebee
gravity is of particular interest, where spontaneous violation of
Lorentz symmetry enters through a nonzero vacuum expectation value
of a bumblebee field when an appropriate potential is brought to
action.

Lorentz symmetry is a fundamental symmetry required for the
formulation of any physically viable theories. Of course, Lorentz
symmetry has had a big impact on the formulation of the Standard
model of particle physics as well as Einstein's general theory of
relativity. The majority of the known physical occurrences in the
universe within the achievable energy range may be satisfactorily
explained by these two major theories. However, evidence from
high-energy cosmic rays \cite{COSMO1, COSMO2} and recent advances
in unified gauge theories suggest that Lorentz symmetry may
spontaneously break in physics at a higher energy scale.
Furthermore, recent studies indicate that some signals related to
Lorentz violation may potentially manifest at lower energy scales,
enabling the discovery of their corresponding consequences in
experiments \cite{SAMUEL}. It has been observed over time that
plenty of theories, including loop quantum gravity, the standard
model extension \cite{SAMUEL, COST1, COST2}, string theory
\cite{COST3}, and others, accommodate the Lorentz violation
scenario. Furthermore, it is expected that studies on Lorentz
symmetry violations would lead to a deeper understanding of
nature.

One of the salient and sound theories that contain Lorentz
violation is the so-called Einstein-bumblebee gravity
\cite{COST4}. In this paradigm, a nonzero vacuum expectation value
of a bumblebee vector field renders possible spontaneous Lorentz
symmetry violation. The publications \cite{BHULUM, BERTO, BAIL,
BHUL, KOSTJT, SEIF, MALUF, GUIO, ESCO, COSMOLOGY, FANG, MARIZ,
ADS, KHODADI, REYER} contain a great deal of research on the
related consequences of the Lorentz violation in black hole
physics and cosmology. In \cite{CASANA}, Casana et al. obtained
the first black hole. solution for such an effective theory,
dubbed Einstein-bumblebee gravity. The gravity of a static,
spherically symmetric, and neutral black hole is precisely
interpreted by this solution. The effect of Lorentz violation on
Hawking radiation has also been studied in \cite{KANZI}. Moreover,
additional spherically symmetric black hole solutions, such as the
traversable wormhole solution \cite{SAKIL}, the global monopole
\cite{OVGUN}, the cosmological constant \cite{JUSUFI}, or the
Einstein-Gauss-Bonnet term \cite{DING}, have been found within the
framework of the bumblebee gravity theory. Furthermore, data
concerning the Lorentz violation have been obtained that was
retained in the black hole shadow \cite{DINGCAS, WANG}, the
accretion disc \cite{LUI}, the black hole superradiance
\cite{RLIN}, and the motion of the massive body \cite{ZLI}. Recent
literature has also provided a solution for the rotating black
hole within the Einstein-bumblebee gravity \cite{DINGCAS}.
Additionally, a black hole that mimics a Kerr-Sen black hole has
been established as a solution to the Einstein-bumblebee model in
\cite{ARJ}. The range of the Lorentz symmetry-breaking parameter
for the rotating black hole in the Einstein-bumblebee gravity is
also constrained by the use of quasi-periodic oscillation
frequencies derived from the data obtained from the observations
(GRO J1655-40, XTE J1550-564), and $GRS 1915+105$ \cite{MOTTA,
MILLER, REID}. These investigations are indeed helpful in figuring
out the effects of LSB by virtue of the bumblebee field.

Standard-Model Extension (SME) is a well-known general effective
field framework that is capable of expressing all conceivable
coefficients for Lorentz violation \cite{SME}. The gravitational
sector of SME is established on a Riemann-Cartan manifold. In
addition to the metric, the torsion is taken into account as a
dynamic geometrical quantity. Even though the gravity sector of
SME is defined in a non-Riemannian context; most research has been
conducted in the metric approach to gravity, where the metric is
the only dynamic geometrical field. It is still worthwhile to take
into account a more generic geometrical framework, even if most of
the research focuses on modified theories of gravity using the
usual metric approach. In this context, we mention the induction
of gravitational topological term \cite{JRN} as one of the more
relevant examples to take advantage of theories of gravity in a
Riemann-Cartan background. One additional intriguing
non-Riemannian geometry that has been discussed in the literature
is the Finsler one \cite{BAO}, which has a number of recent
studies \cite{FOSTER, KOSTE, SCHREK, DCOLLA, SCHREK1} connected to
the LSB contained in it. The so-called metric affine (Palatini)
formalism, in which metric and affine connections are assumed to
have independent dynamic geometrical quantities, is the most
convincing generalization of the metric approach. For a discussion
and some intriguing findings within the Palatini approach, please
see the papers \cite{GHIL, GHIL1}, and references. therein.
Although there have been several recent studies, incorporating
bumblebee gravity \cite{ADEL, ADEL1, ADEL2}, LSB in this scenario
has not received much attention in the literature prior to
\cite{AFFINE}. In the very recent instructive and intriguing study
\cite{AFFINE}, this gap has been attempted to fill up in the
literature by determining the first accurate solution for a
specific metric-affine bumblebee gravity model. It is indeed,
different from those which were proposed in \cite{ADEL, ADEL1,
ADEL2}. Additionally, the function of the LSB parameter has been
explored by contrasting the theoretical findings with the observational data
of the light deflection and the computation of the
perihelion advance of Mercury.

Gravitational waves emitted from black holes are of particular
interest. A black hole that emerges as a result of mass collapsing
gravitationally enters into a perturbed state and releases
radiation that comprises a bundle of distinctive frequencies
unconnected to the collapse process. A very recent research
\cite{AFFINE}, already mentioned in the preceding paragraph
presents a static black hole solution derived from non-Remanian
bumblebee theory, in which LSB enters prominently and plays a
crucial role. This intriguing development in \cite{AFFINE} open a
scope to study the impact of LSB on different astrophysical
aspects linked to the black hole, especially when it is subjected
to perturbation. In this context, we carry out scalar and
electromagnetic perturbation using metric-affine bumblebee theory
and study the impact of LSB on the QNMs, greybody bounds, and
the sparsity of Hawking radiation, and the lifetime of a black hole.

The remainder of the paper is structured as follows. A brief
discussion of the vacuum solution for the metric affine bumblebee
model is provided in Section II. Section III is dedicated to the
computation of QNMs and the analysis of the ringdown wave
structure, with a focus on the impact of LSB on both. Section IV
documents the effect of LSB on Hawking radiation spectra and
scarcity. Sec. V illustrates how the LSB affects Hawking
evaporation and the lifetime of black holes. Section VI includes a
discussion and conclusion of the entire paper.
\section{BLACK HOLE SOLUTION IN METRIC-AFFINE TRACELESS BUMBLEBEE MODEL}
The metric-affine (Palatini) formalism is widely employed in the
study of modified theories of gravity. It is a generalization of
the metric approach where the metric and the connection are
considered independent dynamic geometrical quantities. In the
manuscript \cite{AFFINE}, authors have considered the traceless
metric-affine bumblebee model and came up with
 a static and spherically symmetric vacuum solution where the Lorentz symmetry
 is spontaneously broken. The action for the model reads \cite{AFFINE}
\begin{eqnarray}
\mathcal{S}_{B}&=&\int d^{4}x\,\sqrt{-g}\left[\frac{1}{2\kappa^2}\left(R(\Gamma)
+\xi\left(B^{\mu}B^{\nu}-\frac{1}{4}B^{2}g^{\mu\nu}\right)R_{\mu\nu}(\Gamma)\right)
-\frac{1}{4}B^{\mu\nu}B_{\mu\nu}-\right.\nonumber\\ &-&\left.V(B^{\mu}B_{\mu}\pm b^{2})\right]+
\int d^{4}x\sqrt{-g}\mathcal{L}_{mat}(g_{\mu\nu},\psi),
\label{bumb}
\end{eqnarray}
where $B_{\mu}$ is the bumblebee field, $s^{\mu\nu}$ is traceless
metric, and $V(B^{\mu}B_{\mu}\pm b^2)$ is the potential that
breaks the Lorentz symmetry
 spontaneously, $b^2=b_{\mu}b^{mu}$ being a real positive constant. It is assumed
 that the potential has a minimum at $B^{\mu}B_{\mu}\pm b^2=0$ and $V'(b_{mu}b^{mu})=0$
  to ensure $U(1)$ symmetry breaking, where the bumblebee field acquires a nonzero
  vacuum expectation value, $<B_{\mu}>=b_{\mu}$. With the additional assumption that
  the minimum value of the potential is zero, authors in \cite{AFFINE}, after some
  algebraic steps obtained the following static and spherically symmetric metric
\begin{equation}
ds^2=-\frac{\left(1-\frac{2M}{r}\right)}{\sqrt{\left(1+\frac{3\alpha}{4}\right)
\left(1-\frac{\alpha}{4}\right)}}dt^2+\frac{dr^2}{\left(1-\frac{2M}{r}\right)}
\sqrt{\frac{\left(1+\frac{3\alpha}{4}\right)}{\left(1-\frac{\alpha}{4}\right)^3}}
+r^{2}\left(d\theta^2 +\sin^{2}{\theta}d\phi^2\right),
\label{metric}
\end{equation}
where $\alpha$ is the Lorentz-violating parameter. In the limit
$\alpha\rightarrow 0$, the Lorentz symmetry breaking (LSB) metric
[\ref{metric}] reduces to the Schwarzschild metric. The the
Kretschmann scalar invariant corresponding to this metric is
\begin{eqnarray}
 K &= &R_{\lambda\eta\mu\nu}R^{\lambda\eta\mu\nu}\nonumber \\
 &=&\frac{1}
{r^{6}(4+3\alpha)^{3/2}}
[48 \alpha Mr\sqrt {4+3\alpha}+32M\alpha r\sqrt {4-\alpha}\nonumber \\
\nonumber
 &-& 12 M{\alpha}^{2}r\sqrt {4-\alpha}+32 {r}^{2}\sqrt {4+
3\alpha}+192{M}^{2}\sqrt {4+ 3\alpha}\nonumber \\
 &-& 32{r}^{2}\sqrt {4-\alpha}-16{r}^{2} \alpha\sqrt {4-\alpha}
-12{\alpha}^{2}Mr\sqrt {4+3\alpha} \nonumber \\
&+& 6 {r}^{2}{\alpha}^{2}\sqrt {4-\alpha}+64Mr\sqrt {4-\alpha}-144
\alpha{M}^{2} \sqrt {4+3\alpha}\nonumber \\
&-& 3{M}^{2}{\alpha}^{3}\sqrt {4+3 \alpha}+36
{M}^{2}{\alpha}^{2}\sqrt {4+ 3\alpha}+3 {\alpha}^{2}
{r}^{2}\sqrt {4+ 3\alpha}\nonumber \\
&+&{\alpha}^{3}Mr\sqrt {4+3\alpha}-64Mr\sqrt
{4+3\alpha}-\frac{1}{4}{\alpha}^{3}{r}^{2}\sqrt {4+3\alpha}].
\label{KREST}
\end{eqnarray}
The equation of Kretschmann scalar invariant \ref{KREST} makes it
clear that the effects of LSB, indicated by the parameter
$\alpha$, cannot be fully absorbed by a simple re-scaling of
coordinates. The anticipated standard result corresponds to the
Schwarzschild metric $K_{SC}=\frac{48 M^2}{r^6}$ when $\alpha=0$
is set.

\section{Quasinormal modes and ringdown waveform of the LSB black hole}
In this section, the impact of LSB on the QNMs  due to scalar and
electromagnetic perturbations are studied in detail for the LSB metric
 [\ref{metric}]. The computation of  QNMs is done following the
 important works \cite{iyer, iyer1, konoplya1}
 To study the QNMs in the background of the LSB black hole, we transform
 the relevant equation of the considered field to a Schr$\ddot{o}$dinger-like equation.
  We consider the Klein-Gordon equation for the scalar field and the Maxwell equations for
  the electromagnetic field. For the massless scalar field, we have
\begin{eqnarray}
\frac{1}{\sqrt{-g}}{\partial_\mu}(\sqrt{-g}g^{\mu\nu} \partial_{\nu}\psi) =0,
\label{scalar}
\end{eqnarray}
and for the electromagnetic field, we have
\begin{equation}
\frac{1}{\sqrt{-g}}{\partial_\nu }(F_{\rho\sigma}g^{\rho\mu}g^{\sigma\nu}\sqrt{-g})=0,
\label{em}
\end{equation}
where $ F_{\rho\sigma}={\partial_\rho}A^\sigma-{\partial_\sigma}A^\rho $, $A_\nu$ being
electromagnetic four-potential.
We introduce the tortoise coordinate defined by:
\begin{eqnarray}
\frac{\text{d}r_*}{\text{d}r}=\sqrt{|g_{tt}^{-1}|g_{rr}}.
\label{tortoise}
\end{eqnarray}
With the help of the tortoise coordinate, Eqs.(\ref{scalar}) and (\ref{em}) transform
into the following Schr$\ddot{o}$dinger-like form
\begin{equation}
-\frac{\text{d}^2\phi}{\text{d}{r^2_*}}+V_{\text{eff}}(r) \phi=\omega ^{2}\phi,
\label{schrodinger}
\end{equation}
where the effective potential is given by
\begin{eqnarray}
V_{\text{eff}}(r)&=&|g_{tt}|\left(\frac{\ell(1+\ell)}{r^2}+\frac{1-s^2}{r\sqrt{|g_{tt}|g_{rr}}}
\frac{\text{d}}{\text{d}r}\sqrt{|g_{tt}|g_{rr}^{-1}}\right)\\\nonumber
&=&-\frac{2 (2 M-r) \left(2 \sqrt{3 \alpha +4} \ell (\ell+1) r-(4-\alpha )^{3/2} M \left(s^2-1\right)\right)}{\sqrt{3 \alpha +4} \sqrt{-3 \alpha ^2+8 \alpha +16} r^4}.
\label{vtotal}
\end{eqnarray}
Here, $\ell$ is the angular momentum and $s$ is the spin. The effective potential for the
scalar perturbation is obtained with $s=0$, and $s=1$ yields the
effective potential for the electromagnetic perturbation. Since
the QNMs depend on the nature of the effective potential
[\ref{vtotal}], we illustrate the qualitative nature
 of variation of the effective potential for various scenarios.
\begin{figure}[H]
\centering
\subfigure[]{
\label{vfig1}
\includegraphics[width=0.4\columnwidth]{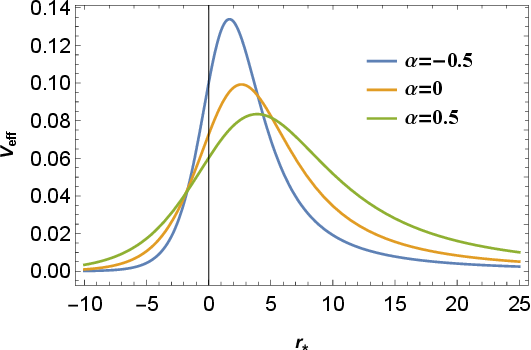}
}
\subfigure[]{
\label{vfig2}
\includegraphics[width=0.4\columnwidth]{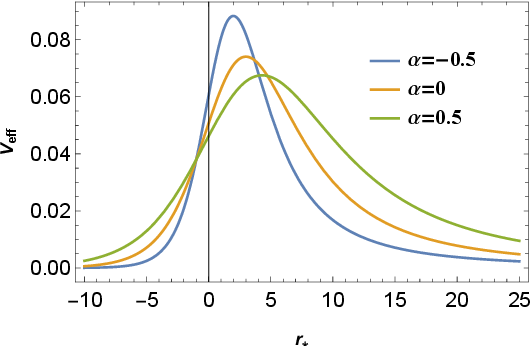}
}
\subfigure[]{
\label{vfig3}
\includegraphics[width=0.4\columnwidth]{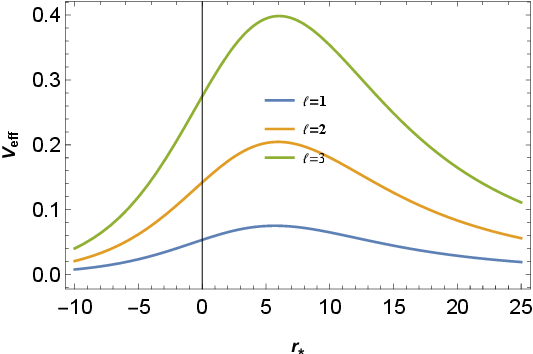}
}
\subfigure[]{
\label{vfig4}
\includegraphics[width=0.4\columnwidth]{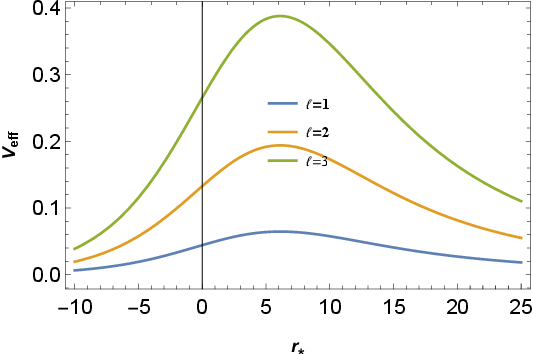}
} \caption{Variation of effective potential with respect to
tortoise coordinate $r_*$. The upper ones are for various values
of $\alpha$ with $\ell=1$ and the lower ones are for various
values of angular momentum $\ell$ with $\alpha=1.0$. The left ones
are for scalar perturbations and the right ones are for
electromagnetic perturbations.} \label{vfig}
\end{figure}
The LSB parameter $\alpha$ has a considerable impact on the
effective potential, as seen in Fig. [\ref{vfig}]. As a
consequence, QNMs will also experience a significant impact that
will be demonstrated later in this section.  The LSB parameter
affects both the peak as well as the position of the peak of the
potential. As the parameter $\alpha$ increases, the peak of the
potential moves to the right and diminishes in value.

Next, we employ the $6th$-order WKB method to obtain QNMs due to
scalar and electromagnetic perturbation. To compute QNMs, the use of
WKB method which was first conducted by Schutz and Will
\cite{schutz}.
 It was later extended to higher orders in the subsequent
 works \cite{iyer, iyer1, konoplya1, konoplya2}.
The $6th$-order WKB method yields the following expression of
quasinormal frequencies:
\begin{equation}
\frac{\text{i}(\omega ^{2}-V_{0})}{\sqrt{-2V_{0}^{''}}}-\sum ^{6}_{\text{i}=2}\Omega_\text{i}=n+\frac{1}{2},
\label{WKB}
\end{equation}
where $V_{0}$ and $V''_{0}$ represent  respectively the height of
the effective potential and the second derivative of it at its
maxima with respect to the tortoise coordinate and
$\Omega_\text{i}$ are the correction terms  according to the
notations followed in \cite{schutz, iyer, iyer1, konoplya1}. To
illustrate the impact of the LSB parameter on the QNMs of the
black
 hole under consideration, we tabulate numerical values of the quasinormal frequencies
 due to
  scalar and electromagnetic perturbations for various values of angular momentum $\ell$
  and parameter $\alpha$. For the overtone number $n=0$, we tabulate quasinormal frequencies
  for scalar
in Table [\ref{QNMS}], and the
   quasinormal frequencies for electromagnetic perturbation are tabulated in Table [\ref{QNMEM}].
   The error associated with the
   $6th$ order WKB method is also provided with the help of the equation given below:
\begin{equation}
\Delta_6=\frac{|\omega_7 -\omega_5|}{2},
\end{equation}
where $\omega_5$ and $\omega_7$ are quasinormal frequencies
obtained using $5th$ order and $7th$ order terms respectively in
the mathematical expression (series) QNMs obtained using the WKB
method.
\begin{table}[!htp]
\centering \caption{Quasinormal frequencies for scalar
perturbation with $n=0$.}
\setlength{\tabcolsep}{1mm}
\begin{tabular}{|c|c|c|c|c|c|c|}
\hline
$\alpha $ & $\ell=1$ & $\Delta _6$ & $\ell$=2 & $\Delta _6$ & $\ell$=3 & $\Delta _6$ \\
\hline
 -0.5 & 0.316497\, -0.130423 i & 0.00030487 & 0.526068\, -0.129494 i & 0.000033815 & 0.736024\, -0.129291 i & $5.28627\times 10^{-6}$ \\
 -0.25 & 0.302284\, -0.111493 i & 0.000171713 & 0.500772\, -0.110525 i & 0.0000163295 & 0.699948\, -0.110283 i & $2.54684\times 10^{-6}$ \\
 0. & 0.29291\, -0.0977616 i & 0.0000985776 & 0.483642\, -0.0967661 i & $8.50954\times 10^{-6}$ & 0.675366\, -0.0965006 i &
  $1.33385\times 10^{-6}$ \\
 0.25 & 0.286619\, -0.0870839 i & 0.0000577862 & 0.471729\, -0.08607 i & $4.68775\times 10^{-6}$ & 0.658126\, -0.0857906 i &
   $7.35039\times 10^{-7}$  \\
 0.5 & 0.282529\, -0.0783729 i & 0.0000337063 & 0.463522\, -0.0773494 i & $2.64843\times 10^{-6}$ & 0.646094\, -0.0770625 i & $4.17979\times 10^{-7}$ \\
\hline
\end{tabular}
\label{QNMS}
\end{table}
\begin{table}[!htp]
\centering \caption{Quasinormal frequencies for electromagnetic
perturbation with $n=0$.}
\setlength{\tabcolsep}{1mm}
\begin{tabular}{|c|c|c|c|c|c|c|}
\hline
$\alpha $ & $\ell=1$ & $\Delta _6$ & $\ell$=2 & $\Delta _6$ & $\ell$=3 & $\Delta _6$ \\
\hline
-0.5 & 0.257511\, -0.121984 i & 0.000607021 & 0.49203\, -0.126647 i & 0.0000285133 & 0.711934\, -0.127869 i & $4.47326\times 10^{-6}$ \\
 -0.25 & 0.252083\, -0.105091 i & 0.000287745 & 0.471644\, -0.108353 i & 0.0000135937 & 0.679313\, -0.109193 i & $2.16552\times 10^{-6}$ \\
 0. & 0.248191\, -0.092637 i & 0.000144615 & 0.457593\, -0.0950112 i & $7.02456\times 10^{-6}$ & 0.656898\, -0.0956171 i &
   $1.13678\times 10^{-6}$ \\
 0.25 & 0.245539\, -0.0828292 i & 0.0000768655 & 0.447725\, -0.0845992 i & $3.82553\times 10^{-6}$ & 0.641097\, -0.0850482 i &
   $6.27721\times 10^{-7}$ \\
 0.5 & 0.243917\, -0.0747463 i & 0.0000415175 & 0.440902\, -0.0760844 i & $2.17153\times 10^{-6}$ & 0.630037\, -0.0764224 i &
   $3.58897\times 10^{-7}$ \\
\hline
\end{tabular}
\label{QNMEM}
\end{table}
The impact of LSB on quasinormal frequency is distinctly visible
from the above tables. We can infer from Tables [\ref{QNMS},
\ref{QNMEM}] that the real part of quasinormal frequencies that
corresponds to the frequency of the gravitational waves decreases with an increase in $\alpha$ for both types of
perturbations. On the other hand, the imaginary part of the
quasinormal frequencies show the opposite effect. This implies that the
frequency of gravitational waves as well as the damping rate
decreases with $\alpha$. The variation
of the decay rate with the angular momentum, however, is marginal. However, the impact of $\ell$ on decay rate for scalar and electromagnetic perturbations are opposite in nature. For scalar perturbation, the decay rate decreases with $\ell$ and for electromagnetic perturbation, the decay rate increases with $\ell$. Additionally, we observe that the error follows the same pattern
as followed by the decay rate or the frequency. Tables
[\ref{QNMS}, \ref{QNMEM}] have shown quantitative variation of
quasinormal frequencies. For completeness, we illustrate the
qualitative nature of variation through a graphical presentation of
quasinormal frequencies for various situations. A comparison of
real and imaginary parts of quasinormal frequencies due to scalar
and electromagnetic perturbations are also presented through
plots.
\begin{figure}[H]
\centering
\subfigure[]{
\label{qnmimfig1}
\includegraphics[width=0.4\columnwidth]{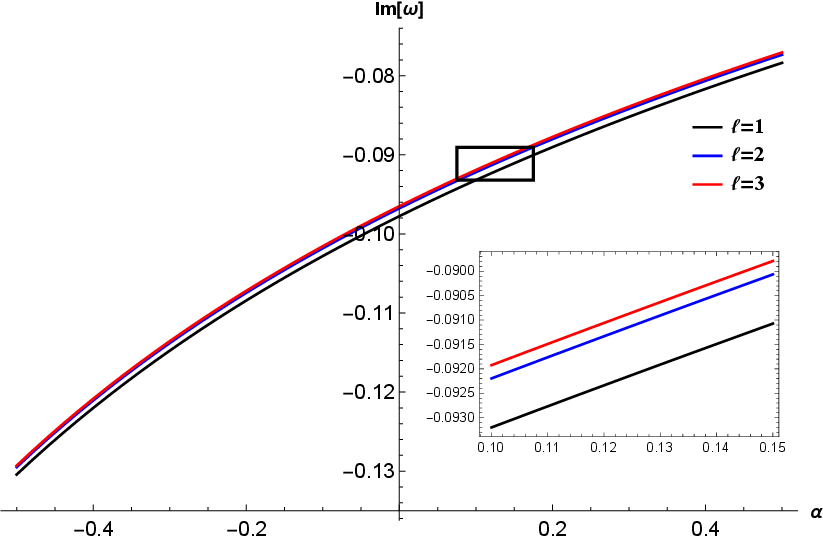}
}
\subfigure[]{
\label{qnmimfig2}
\includegraphics[width=0.4\columnwidth]{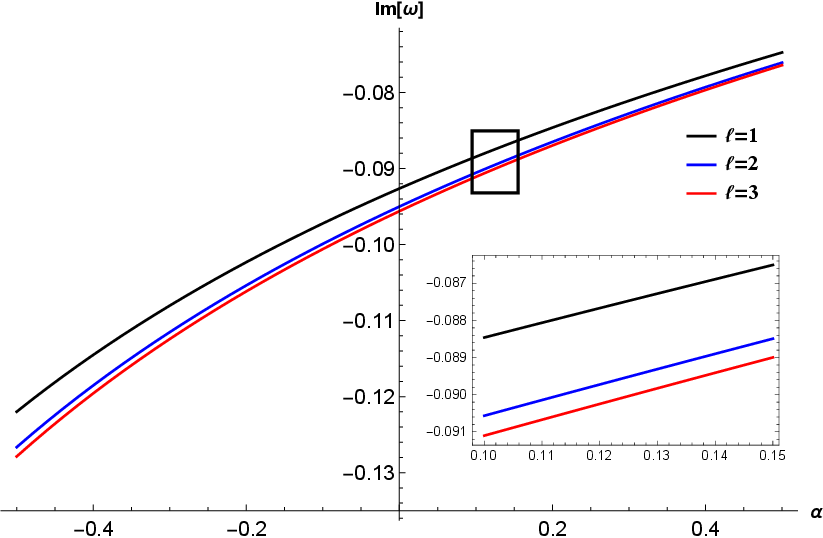}
}
\caption{It gives the variation of the imaginary part of the quasinormal frequency with
respect to $\alpha$ for various values of $\ell$. The left one is for the scalar field and
the right one is for the electromagnetic field.}
\label{qnmimfig}
\end{figure}
\begin{figure}[H]
\centering
\subfigure[]{
\label{qnmrefig1}
\includegraphics[width=0.4\columnwidth]{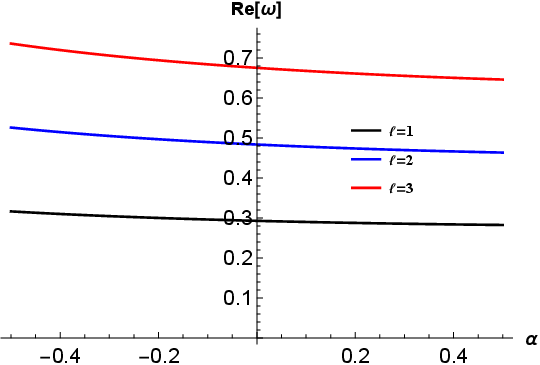}
}
\subfigure[]{
\label{qnmrefig2}
\includegraphics[width=0.4\columnwidth]{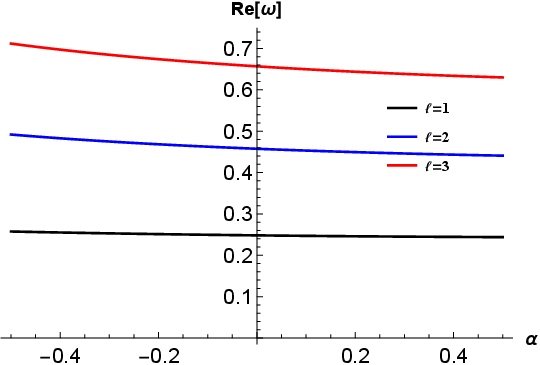}
}
\caption{It gives the variation of the real part of the quasinormal frequency with
respect to $\alpha$ for various values of $\ell$. The left one is for the scalar field
and the right one is for the electromagnetic field.}
\label{qnmrefig}
\end{figure}
\begin{figure}[H]
\centering
\subfigure[]{
\label{qnmrefig1}
\includegraphics[width=0.4\columnwidth]{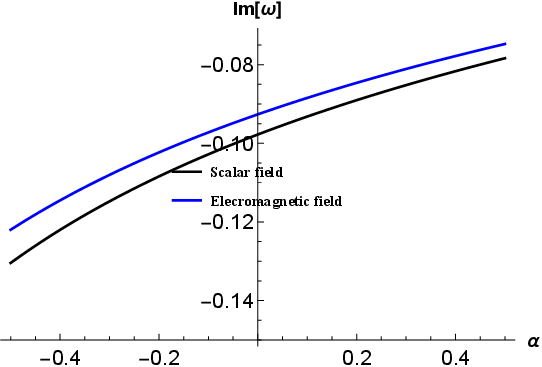}
}
\subfigure[]{
\label{qnmrefig2}
\includegraphics[width=0.4\columnwidth]{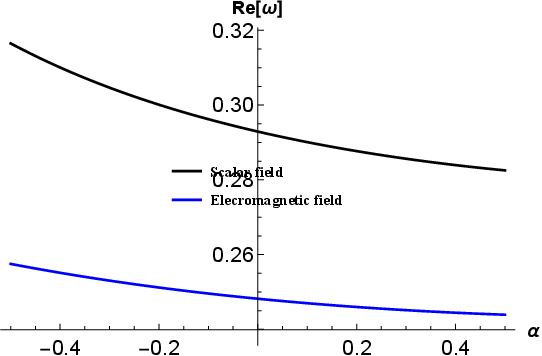}
}
\caption{The Left one gives the variation of the imaginary part of the quasinormal
frequency with respect to $\alpha$ for scalar and electromagnetic fields and the right
one gives that for the real part. Here, we have taken $\ell=1$.}
\label{qnmimrefig}
\end{figure}
Fig. [\ref{qnmimfig}] and Fig. [\ref{qnmrefig}] reinforce what we have already known
from Tables [\ref{QNMS}, \ref{QNMEM}]. Fig. [\ref{qnmimrefig}] shows that the frequency
of the gravitational wave as well as the decay rate is larger for scalar perturbation.
 The convergence of the WKB method for various values of $(n,\ell)$ pair is graphically
 illustrated below.
\begin{figure}[H]
\centering
\subfigure[]{
\label{qnmorderfig3}
\includegraphics[width=0.4\columnwidth]{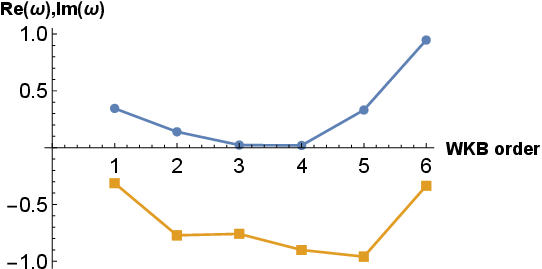}
}
\subfigure[]{
\label{qnmorderfig4}
\includegraphics[width=0.4\columnwidth]{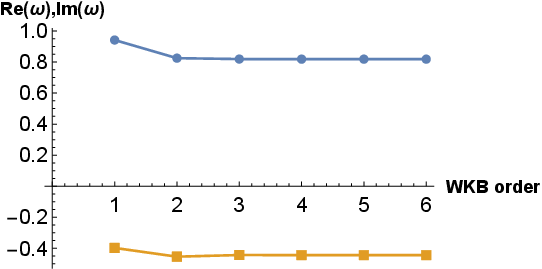}
} \caption{Variation of the real and imaginary parts of
quasinormal frequencies with respect to WKB order for various
values of the $(n,\ell)$ pair is shown. The left one is for the
(3,0) pair and the right one is for the (2,4) pair. In each plot,
the blue line is for the real part, and the orange line is for
the imaginary part of the QNMs. Here, we have taken $\alpha=0.2$.}
\label{qnmorderfig}
\end{figure}
Fluctuation of quasinormal frequencies even for higher order for
the pair $(3,0)$ as shown in Fig. [\ref{qnmorderfig}] that
confirms the conclusion drawn in the manuscript \cite{kd} where
 it is observed that the WKB approximation is reliable when the angular momentum is high and
  the overtone number is low.

The study of the time evolution of perturbation profiles provides
another avenue to observe the impact of the Lorentz violation. To
this end, the time domain integration method formulated by
 Gundlach et al. in their article \cite{gundlach1} is employed here to numerically solve the
 time-dependent wave equation. We use the initial conditions $\psi(r_*,t) =
 \exp \left[ -\dfrac{(r_*-\hat{r}_{*})^2}{2\sigma^2} \right]$ and $\psi(r_*,t)\vert_{t<0} = 0$
 with $r_*=5$, $\hat{r}_*=0.4$. While choosing the values
 of $\Delta t$ and $\Delta r_{*}$,
 we take into account the Von Neumann stability
 condition, $\frac{\Delta t}{\Delta r_*} < 1$.

\begin{figure}[H]
\centering
\subfigure[]{
\label{ringingscalar1}
\includegraphics[width=0.45\columnwidth]{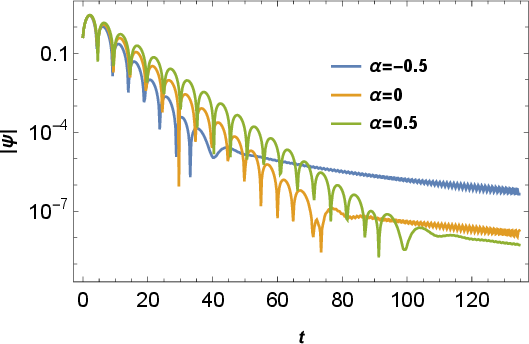}
}
\subfigure[]{
\label{ringingscalar2}
\includegraphics[width=0.45\columnwidth]{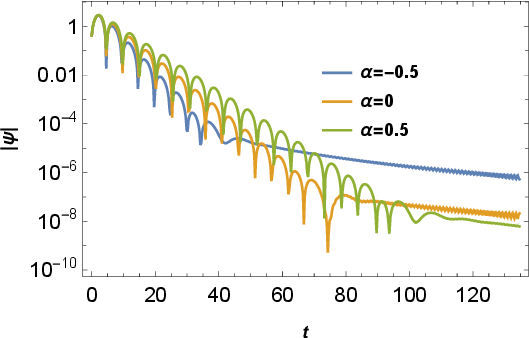}
}
\caption{Time domain profile for various values of $\alpha$. The left one is for scalar
perturbation and the right one is for electromagnetic perturbation. Here, we have taken $\ell=3$.}
\label{ringingalpha}
\end{figure}

\begin{figure}[H]
\centering
\subfigure[]{
\label{ringingem1}
\includegraphics[width=0.45\columnwidth]{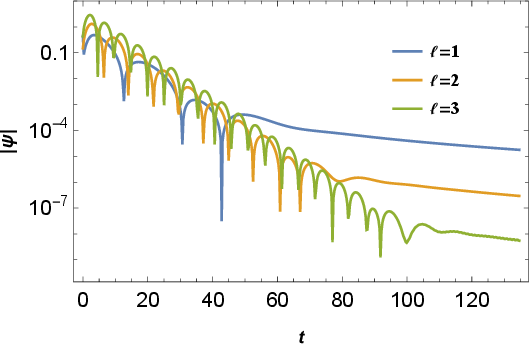}
}
\subfigure[]{
\label{ringingem2}
\includegraphics[width=0.45\columnwidth]{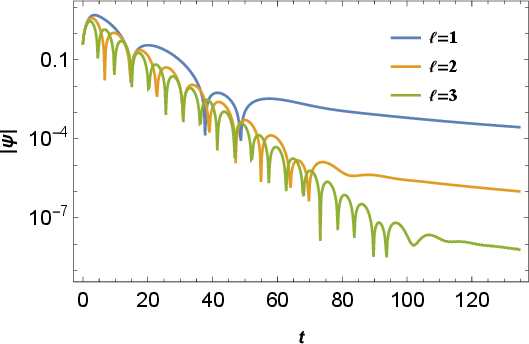}
}
\caption{Time domain profile for various values of $\ell$. The left one is for scalar
perturbation and the right one is for electromagnetic perturbation. Here, we have taken $\alpha=0.5$.}
\label{ringingl}
\end{figure}

The dependence of the ringdown waveform on the LSB parameter
$\alpha$ is graphically illustrated in the Fig.
[\ref{ringingalpha}] and the ringdown waveforms for various values
of $\ell$ are shown in the Fig. [\ref{ringingl}]. These figures
indicate the effect of LSB time profiles of perturbations and
confirm our conclusions drawn from Tables [\ref{QNMS},
\ref{QNMEM}].

\section{Spectrum and sparsity of Hawking radiation}
\begin{figure}[H]
\centering \subfigure[]{\label{hw1}
\includegraphics[width=0.4\columnwidth]{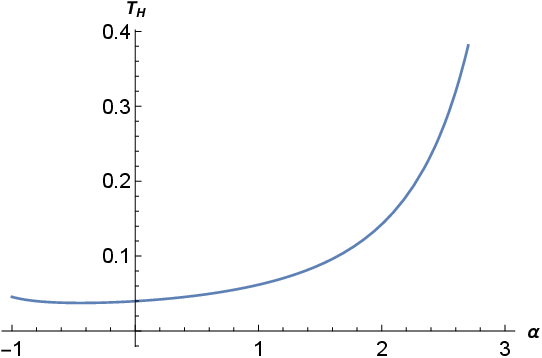}}
\caption{Variation of Hawking temperature with respect to $\alpha$.}
\label{hwfig}
\end{figure}

Here, we investigate the effect of the LSB parameter on the spectrum and the sparsity of
the Hawking radiation. For this purpose, we first need to calculate the Hawking temperature
and the greybody bounds. The Hawking temperature is given by
\begin{equation}
T_H=\frac{1}{4\pi \sqrt{-g_{tt}g_{rr}}}\frac{dg_{tt}}{dr}|_{r=r_h},
\end{equation}
where $r_h$ is the position of the event horizon which is $2M$ in
our case. Putting metric coefficients from [\ref{metric}] in the
above equation, we get
\begin{equation}
T_H=\frac{1}{8 \pi \sqrt{\left(1-\frac{\alpha}{4}\right)^5 \left(\frac{3 \alpha}{4}+1\right)} M}.
\label{hawking}
\end{equation}
The above expression reduces to that for the Schwarzschild case in the limit $\alpha\rightarrow 0$.
 To have a qualitative idea of the variation of the Hawking temperature with $\alpha$, we plot
 $T_H$ with respect to $\alpha$ in Fig. [\ref{hwfig}]. It shows that the Hawking temperature
 initially decreases with $\alpha$ reaching a minimum value at $\alpha=-0.444445$ and then
 starts increasing with $\alpha$.
The next ingredient required to study the spectrum and sparsity is
the greybody bound given by \cite{GB, GB1, GB2,ac2020}
\begin{equation}
T\geq sech^{2}(\frac{1}{2\omega}\int_{-\infty}^{\infty}|V_{\text{eff}}(r_*)|dr_*).\label{grey}
\end{equation}
The analytical expression for the greybody bound of the black hole
under consideration is provided in the article \cite{affine1}.
Now, with the requisite information in hand, we are in a position
to delve into the central topic of this section. The expression
for the total power emitted by a black hole in the form of Hawking
radiation is \cite{yg2017, fg2016}
\begin{equation}
\frac{dE(\omega)}{dt}\equiv P_{tot} = \sum_\ell T (\omega)\frac{\omega}{e^{\omega/T_{H}}-1}
\hat{k} \cdot \hat{n}~ \frac{d^3 k ~dA}{(2\pi)^3},
\end{equation}
where $dA$ is the surface element, $\hat{n}$ is unit normal to $dA$, and $T$ is the greybody
factor given by Eq. (\ref{gb}). For massless particles $|k|=\omega$ which reduces above expression to
\begin{equation}\label{ptot}
P_{tot}=\sum_\ell \int_{0}^{\infty} P_\ell\left(\omega\right) d\omega.
\end{equation}
Here, $P_{\ell}$ is the power spectrum in the $\ell th$ mode given by
\begin{equation}\label{pl}
P_\ell\left(\omega\right)=\frac{A}{8\pi^2}T(\omega)\frac{\omega^3}{e^{\omega/T_{H}}-1}.
\end{equation}
Here, $A$ is a multiple of the horizon area. We take it to be equal to the horizon area as
it will not affect the qualitative result \cite{yg2017}. To study the qualitative nature of
variation of the power spectrum with respect to the LSB parameter $\alpha$, we plot $P_{\ell}(\omega)$
against $\omega$ for various values of $\alpha$.
\begin{figure}[H]
\centering
\subfigure[]{
\label{pls}
\includegraphics[width=0.4\columnwidth]{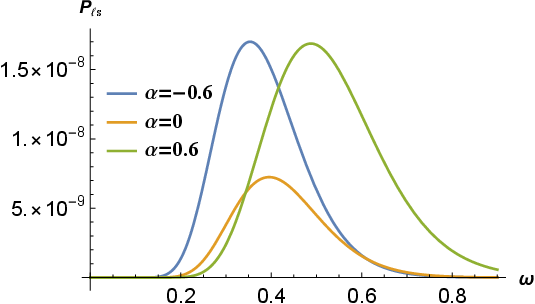}
}
\subfigure[]{
\label{plem}
\includegraphics[width=0.4\columnwidth]{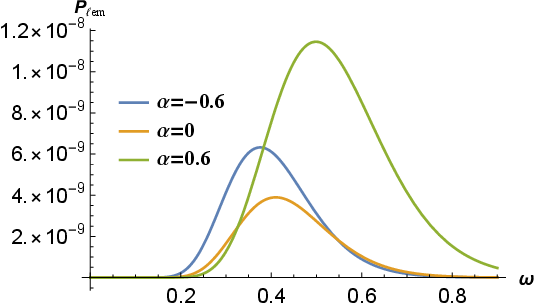}
}
\caption{Power spectrum of the black hole for various values of $\alpha$. The left one is
for scalar perturbation and the right one is for electromagnetic perturbation. Here,
 we have taken $\ell=2$.}
\label{pl}
\end{figure}
An interesting observation we can make from Fig. [\ref{pl}] is
that for both perturbations, $P_{\ell}(\omega)$ increases for
non-zero values of the LSB parameter, whereas the peak shifts
towards left for negative values of $\alpha$ and shifts towards
right for positive values of $\alpha$.

The sparsity of the Hawking radiation can be quantitatively
measured with the help of a dimensionless quantity $\eta$ defined
by \cite{yg2017, fg2016, ac2020, sh2016, sh2015}
\begin{equation}
\eta=\frac{\tau_{gap}}{\tau_{emission}}.
\label{eta}
\end{equation}
Here, $\tau_{gap}$ and $\tau_{emission}$ are the average time gap
between two successive radiation quanta and the time taken by a
radiation quantum for emission, respectively. They are defined as
follows:
\begin{eqnarray}\label{tgap}
\tau_{gap}=\frac{\omega_{max}}{P_{tot}} \quad \text{and} \quad \tau_{emission}
 \geq \tau_{localization}=\frac{2 \pi}{\omega_{max}},
\end{eqnarray}
where $\tau_{localization}$ corresponds to the time period of the emitted wave of
 frequency $\omega_{max}$. The significance of the quantity $\eta$ lies in the fact
  that it provides insight into the continuous or discontinuous nature of flow of the Hawking radiation. For $\eta\ll1$, we have a continuous flow of radiation, whereas large values of $\eta$ signify a discontinuous flow of radiation. Tables [\ref{sparsescalar}, \ref{sparseem}] provide numerical values of $\omega_{max}$, $P_{max}$, $P_{tot}$, and $\eta$ for scalar and electromagnetic perturbations, respectively. A common trend we can observe from the tables is that $P_{max}$ and $P_{\text{tot}}$ initially decreases with $\alpha$ and then starts increasing. The sparsity, on the other hand, initially increases and then decreases with an increase in the LSB parameter. It clearly shows the significant impact of the Lorentz violation effect on Hawking radiation.
\begin{table}[!htp]
\centering
\caption{Numerical values of $\omega_{max}$, $P_{max}$, $P_{tot}$, and $\eta$ for
 scalar perturbation for various values of $\alpha$ for $\ell=1$ mode.}
\setlength{\tabcolsep}{-.2mm}
\begin{tabular}{|c|c|c|c|c|c|c|c|c|c|}
\hline
$\alpha$ & -1. & -0.5 & 0. & 0.5 & 1. & 1.5 & 2.0 & 2.5 & 3.0\\
\hline
$ \omega _{max }$ & 0.171431 & 0.201873 & 0.235445 & 0.286453 & 0.365994 & 0.497487 & 0.735941 & 1.24294 & 2.69832\\
\hline
$ P_{max }$ & 0.00005 & 5.49255$\times 10^{-6}$ & 3.42190$\times 10^{-6}$ & 4.90236$\times 10^{-6}$ & 0.00001 & 0.00005 & 0.00038 & 0.00513 & 0.17181  \\
\hline
$ P_{\text{tot}}$ & 9.6149$\times 10^{-6}$ & 9.81224$\times 10^{-7}$ & 6.71889$\times 10^{-7}$ & 1.15588$\times 10^{-6}$ & 3.84193$\times 10^{-6}$ & 2.30024$\times 10^{-5}$ & 2.57345$\times 10^{-4}$ &
   6.32755$\times 10^{-3}$ & 5.32169$\times 10^{-1}$ \\
\hline
$ \eta$ & 486.468 & 6610.08 & 13131.1 & 11298.3 & 5549.05 & 1712.42 & 334.958 & 38.8585 & 2.1775 \\
\hline
\end{tabular}
\label{sparsescalar}
\end{table}
\begin{table}[!htp]
\centering
\caption{Values of $\omega_{max}$, $P_{max}$, $P_{tot}$, and $\eta$ for electromagnetic
perturbation for various values of $\alpha$ for $\ell=1$ mode.}
\setlength{\tabcolsep}{-.2mm}
\begin{tabular}{|c|c|c|c|c|c|c|c|c|c|}
\hline
$ \alpha$ &  -1. & -0.5 & 0. & 0.5 & 1. & 1.5 & 2.0 & 2.5 & 3.0\\
\hline
$ \omega _{\max }$ &  0.262231 & 0.243064 & 0.263861 & 0.308454 & 0.385231 & 0.514836 & 0.752186 & 1.25849 & 2.71274 \\
\hline
$ P_{\max }$ & 6.28359$\times 10^{-6}$ & 1.37354$\times 10^{-6}$ & 1.33432$\times 10^{-6}$ & 2.60945$\times 10^{-6}$ & 8.39995$\times 10^{-6}$ & 0.00004 & 0.00034 & 0.00484 & 0.16908  \\
\hline
$P_{tot}$ & 1.39766$\times 10^{-6}$ & 2.63489$\times 10^{-7}$ & 2.74252$\times 10^{-7}$ & 6.34356$\times 10^{-7}$ & 2.61403$\times 10^{-6}$ & 1.82744$\times 10^{-5}$ & 2.27702$\times 10^{-4}$ &
   5.99975$\times 10^{-3}$ & 5.2402$\times 10^{-1}$ \\
\hline
$ \eta$ & 7830.45 & 35686.1 & 40403.7 & 23870.9 & 9035.49 & 2308.42 & 395.461 & 42.013 & 2.23506\\
\hline
\end{tabular}
\label{sparseem}
\end{table}
\section{Hawking evaporation and the lifetime of LSB black holes}
Black holes emit radiation that causes a reduction in the black
hole mass and gives rise to Hawking evaporation. We, in this
section, intend to obtain the 'lifetime', $\tau$, of LSB black
holes. With the help of the Stefan-Boltzmann radiation law, the
rate of energy loss is given by
\begin{equation}
\frac{dM}{dt}=-\gamma M^{2} T_{H}^4,
\end{equation}
where $\gamma$ is a constant that is related to the greybody factor and the radiation
 constant, $T_{H}$ is the Hawking temperature given by Eq. [\ref{hawking}] \cite{ong}.
 The above equation, with the help of Eq. [\ref{hawking}], yields
 \begin{equation}
\frac{dM}{dt}=-\frac{\gamma}{4096 \pi ^4 \left(1-\frac{\alpha}{4}\right)^{10}
 \left(\frac{3 \alpha}{4}+1\right)^2 M^2}.
\end{equation}
To obtain the lifetime of LSB black holes, we integrate the above equation from
 the initial mass $M_i$ to $0$. That produces the result:
\begin{equation}
\tau=\frac{4096 \pi ^4 \left(1-\frac{\alpha}{4}\right)^{10}
\left(\frac{3 \alpha}{4}+1\right)^2 M_{i}^3}{3 \gamma}.
\end{equation}
The above result clearly exhibits the dependence of $\tau$ on the
LSB parameter $\alpha$. To analyze the qualitative variation of
$\tau$ with $\alpha$, we plot $\tau$  versus $\alpha$ with
$M_i=\gamma=1$.
\begin{figure}[H]
\centering \subfigure[]{ \label{hw1}
\includegraphics[width=0.4\columnwidth]{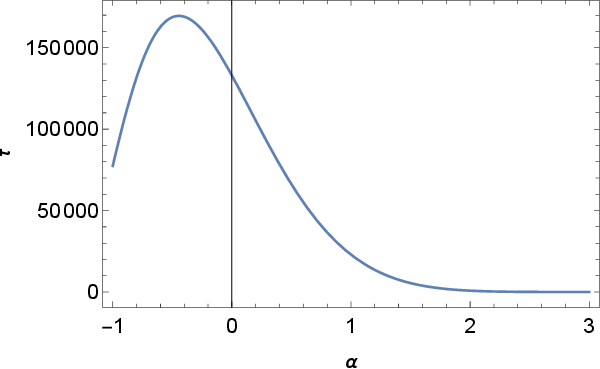}
}
\caption{Variation of $\tau$ with respect to $\alpha$.}
\label{lifetime}
\end{figure}
Figure [\ref{lifetime}] shows us that a black hole's lifetime
first rises with $\alpha$, reaches its highest value at
$\alpha=-0.444444$, and then falls. The maximum value of the
Hawking temperature is found in proximity to the lowest value.
\section{Discussion and conclusions}
Using the static solution provided in \cite{AFFINE} from the
metric-affine bumblebee model, in this work, we have computed the
correction owing to the LSB effects on the QNMs, ringdown
waveforms, Hawking radiation spectra, the sparsity of Hawking
radiation, and the lifetimes of black holes presenting a variety of
graphs and tables. The impact of the LSB on these black
hole-related physical properties have been thoroughly examined.

To illustrate the impact of LSB on QNMs, we have employed the
$6th$-order WKB method. To start with, we provide numerical values
for QNMs for scalar and electromagnetic perturbations in tabular
form, then illustrate graphically how the real and imaginary parts
of QNMs vary with respect to the LSB parameter $\alpha$. Errors
associated with our numerical calculations are also provided.
Additionally, we compare the decay rate and the frequency of
gravitational waves of scalar and electromagnetic perturbations.
Our investigation reveals that the decay rate associated with the
imaginary part of the QNMs and the frequency of gravitational
waves corresponding to the real part of the QNMs initially
decreases with the increase in the LSB parameter $\alpha$, and
then starts increasing. Note that the frequency of the
gravitational wave and decay rate are found to be more sensitive
to negative values of the LSB parameter $\alpha$ than to its
positive values (see Fig. [4]). As a result, the LSB is expected
to be crucially important for maintaining the stability of the
system if receives perturbation from either a scalar or
electromagnetic field. It is expected to hold for other types of
perturbation. We also observe that the real part of QNMs increases
with the angular momentum $\ell$, but the impact of $\ell$ on the
decay rate is marginal. The comparison between scalar and
electromagnetic perturbations reveals that both the decay rate and
the frequency of gravitational waves are a little larger for the
scalar perturbation in comparison to the electromagnetic
perturbation. We have also studied the evolution of perturbation
profiles, which confirm the conclusions we have already drawn from
qualitative and quantitative studies. In Fig. [\ref{qnmorderfig}],
we illustrate the convergence of the WKB. method for various
$(n,\ell)$ pairs. We observe that, for $n < \ell$, QNMs fluctuate
even for higher order.

We then delve into the study of the Hawking spectrum and its
sparsity. In order to accomplish this, we first determine the
Hawking temperature and the greybody bounds for the black hole
under consideration. The variation of the Hawking temperature is
shown in Fig. [\ref{hwfig}]. The plot reveals that the Hawking
temperature initially decreases to a minimum value
at$\alpha=-0.444445$ and then increases with $\alpha$. With the
help of the Hawking temperature and the graybody bounds, we study
the variation of the power spectrum $P_{\ell}(\omega)$ with
respect to the LSB parameter $\alpha$. We can infer from Fig.
[\ref{pl}] that for both types of perturbation $P_{\ell}(\omega)$
increases for non-zero values of the LSB parameter, whereas the
peak shifts towards the left for negative values of $\alpha$, and
for positive values of $\alpha$, it shifts towards the right. We
then tabulate. numerical values of $\omega_{max}$, $P_{max}$,
$P_{tot}$, and $\eta$ for both  scalar and electromagnetic
perturbations. The findings indicate that with the increase
in$\alpha$, there is an initial decline of $P_{max}$, and
$P_{\text{tot}}$, followed by an enhancement. On the other hand,
the sparsity, i.e., the time gap between successive radiation
quanta, initially increases with $\alpha$ and then falls off.
Besides, our study on Hawking evaporation shows that the Lorentz
symmetry violation has a significant impact on the lifetime of
black holes associated with the metric-affine bumblebee model.

In this endeavor, we provide our findings that demonstrate the
significant influence of the LSB on several observables, such as
QNMs, Hawking temperature, sparsity of Hawking radiation, temporal
evaluation of perturbation profiles, and lifetime of black hole.
Our knowledge of modified theories of gravity and its
astrophysical ramifications will be improved by these findings.
Nevertheless, the experiment will establish to what extent a
modified theory of gravity is applicable. Experiments in this
energy regime are, however, sparse. Precise standardization of
this type of theory is therefore currently outside the purview. In
this context, we must mention that in \cite{AFFINE}, we found an
effort to constrain the LSB parameter using low-energy
experimental observation.

\end{document}